\documentclass[aps,twocolumn,prl]{revtex4}
\usepackage[hypertex,linkcolor=red]{hyperref}
\usepackage{exscale}
\usepackage{graphicx}
\usepackage{amsmath}
\usepackage{latexsym}
\usepackage{amsfonts}
\usepackage{amssymb}

\begin{document} 

\textbf{Reply to Comment on ``Regional Versus Global Entanglement in Resonating-Valence-Bond States''}

In the Letter \cite{amader}, we consider \emph{general} resonating-valence-bond (RVB) states on arbitrary lattices 
(please see \cite{amader} for the exact definition considered). 
We prove that the whole state possesses genuine multiparty entanglement, and provide bounds 
on two-particle entanglement. We also have a conjecture regarding the amount of entanglement in large and small portions of the whole
lattice. The proof, bound, and conjecture, are not  considered in the Comment in Ref. \cite{onnoder}.

We then consider two examples: the so-called ``RVB gas'' and ``RVB liquid''. In both cases, for arbitrary lattice-size,
we derive bounds on two-particle entanglement, by several methods, including known methods in condensed matter physics 
(e.g. from Ref. \cite{eita-Tasaki})
and, to our knowledge, hitherto unknown ones that uses quantum information paradigms (specifically,  
``monogamy'' and ``telecloning'').  The case of RVB liquid (for arbitrary lattice size) is not considered in the Comment. 
The case of RVB gas is derived by an independent method in the Comment, and it agrees with our result.

Lastly, we had considered an RVB liquid on a finite-sized lattice (\(4 \times 4\) square lattice) 
and numerically calculated the value of two-particle entanglement. 
(Please note that \emph{all} the other results in our Letter are analytical.) There was an error in the calculation in the second decimal 
place. We have since then redone the calculation. The Authors of the Comment have calculated this quantity by exact diagonalization, and 
have also noticed this error. However, the conclusion of ``no two-particle entanglement'' is still true for the 
RVB liquid on a \(4 \times 4\) square lattice.

Please note that \emph{all} the other results in our Letter are independent of this numerical result.

There exist statements in the Comment on our Letter that are either wrong, or have been written due to 
misunderstanding, and have resulted in a total disarray. To remove the confusion, we enumerate here some of the mistakes in the 
Comment:
\begin{enumerate}
\item[(1)] In the first paragraph of the Comment, the statement\\\\
 ``for the RVB liquid on the square lattice, the calculations and conclusions''
of our Letter ``are incorrect.''\\\\
 is wrong. The calculations and conclusions in Letter, on multiparty and two-party entanglement, 
for lattices of arbitrary size (including 
square ones) are correct. The only mistake is in a numerical calculation (in the second decimal place) 
of two-party entanglement of the \(4 \times 4\) square lattice.
Note however that the conclusion of ``no two-site entanglement'' is still true for the RVB liquid on a \(4 \times 4\) square lattice.
All other results in the paper are independent of this numerical error.

\item[(2)] In the concluding paragraph of the Comment, 
the statement\\\\
 ``the results'' of our Letter ``are incorrect in the other one.''\\\\
 is similarly wrong. Please 
refer to \emph{item (1)} above.
\end{enumerate}

Similar related incorrect statements are scattered in the text of the Comment.

Let us reiterate here that an important motivation for our Letter was to find an independent approach 
for an important and well-known problem. After the approach has been found, one of course needs to compare it with the existing ones, for 
which there is the need of some sort of optimization. One needs to answer questions like
``What is the best bound possible in the new approach?'' The latter has not yet been done, and we wish to pursue it in a later publication.

Dagomir Kaszlikowski\(^1\), Aditi Sen(De)\(^2\), Ujjwal Sen\(^2\), and Vlatko Vedral.\(^{1,3}\)\\
\(^1\) Department of Physics, National University of Singapore, 117542 Singapore, Singapore.\\
\(^2\)ICFO-Institut de Ci\`encies Fot\`oniques, Mediterranean Technology Park,
08860 Castelldefels (Barcelona), Spain.\\
\(^3\)The School of Physics and Astronomy, University
of Leeds, Leeds, LS2 9JT, United Kingdom.

\end{document}